# Room-temperature photonic quantum computing in integrated silicon photonics with germanium-silicon single-photon avalanche diodes

**NEIL NA,**[1,*] **CHOU-YUN HSU,**[1] **ERIK CHEN,**[1] **AND RICHARD SOREF**[2]

[1]*Artilux Inc., Zhubei City, Hsinchu County 30288, Taiwan ROC*
[2]*Department of Engineering, The University of Massachusetts, Boston, Massachusetts 02125, USA*
*\*neil@artiluxtech.com*

**Abstract:** Most, if not all, photonic quantum computing (PQC) relies upon superconducting nanowire single-photon detectors (SNSPDs) based on niobium (Nb) operated at a temperature < 4 K. This paper proposes and analyzes 300 K waveguide-integrated germanium-silicon (GeSi) single-photon avalanche diodes (SPADs) based on the recently demonstrated normal-incidence GeSi SPADs operated at room temperature, and shows that their performance is competitive against that of SNSPDs in a series of metrics for PQC with a reasonable time-gating window to resolve the issue of dark-count rate (DCR). These GeSi SPADs become photon-number-resolving avalanche diodes (PNRADs) by deploying a spatially-multiplexed *M*-fold-waveguide array of *M* GeSi SPADs. Using on-chip waveguided spontaneous four-wave mixing (SFWM) sources and waveguided field-programmable interferometer mesh (FPIM) circuits, together with the high-metric SPADs and PNRADs, high-performance quantum computing at room temperature is predicted for this PQC architecture.

## 1. Introduction

The field of optical quantum information processing, including sub-fields that use photons for quantum computing [1], quantum communication [2], and quantum metrology [3], to name a few, has greatly evolved in the past several decades. Particularly, due to advancement in photonic integrated circuit (PIC) technologies, quantum states represented by the polarization/position/energy etc. degrees-of-freedom of single photons can now be efficiently controlled. This enables the implementation of large-scale photonic quantum computing (PQC) systems, which usually consists of three main building blocks [4]: (1) quantum sources that generates single photons, (2) quantum circuits that manipulates single photons, and (3) quantum detectors that measures single photons. These main three building blocks can be integrally or separately implemented on chips, with fiber-optic interfaces sometimes between the chips as well as sometimes between the chips and free-space optics. Various quantum algorithms that may possess advantages over classical algorithms have been demonstrated, including photon-based (or number-based) computations such as two-photon quantum machine learning [5], Gaussian boson sampling [6], and qubit-based (or gate-based) computations such as prime number factorization [7], variational eigenvalue solving [8].

In the literature, there is a large body of PQC works, both theoretical and experimental, on operating at a cryogenic temperature usually < 4 K mainly because of the essential need to employ on-chip superconducting nanowire single-photon detectors (SNSPDs). It has been universally perceived-and-believed that SNSPDs have unmatched and superior performance metrics, and offer the most practical approach to integrating quantum detectors with PICs. These SNSPD assumptions and reliances are called into question here. In this paper, a new "room-temperature PQC paradigm" is proposed, based on the recently demonstrated normal-incidence germanium-silicon (GeSi) single-photon avalanche photodiode (SPAD) operated at 300 K [9], and we show that it is the key enabler of the proposed room-temperature PQC paradigm when a reasonable time-gating is applied. This paradigm operates at shortwave infrared (SWIR) wavelengths and uses the Si-wafer-based Si-on-insulator (SOI) and Ge-on-Si (GOS) technologies, which have been the standards for silicon photonics (SiPh) to achieve complementary metal-oxide-semiconductor (CMOS) fabrication compatibility. In the following, we shall discuss (1) the general considerations of PQC operated at room temperature, (2) the design and optimization of a proposed waveguide-integrated GeSi SPAD featuring larger than 95 % quantum efficiency (QE) simultaneously at the C and O bands, (3) the analysis on scaling the dark count rate (DCR) of the demonstrated normal-incidence GeSi SPAD to estimate the DCR of the proposed waveguide GeSi SPAD, (4) the considerations and derivations of figures of merit for the single-photon detector (SPD) as well as the photon-number-resolving detector or "number-photon detector" (NPD) used in PQC to benchmark the proposed 300 K waveguide GeSi SPAD with the present-art < 4 K waveguide SNSPD [10-12], and (5) the possibility of performing PQC in SiPh at midwave infrared (MIR) wavelengths using SOI waveguide [13] and GeSn-based detectors [14-17].

**2. General considerations of PQC at room temperature**

Fig. 1 shows the proposed room-temperature PQC paradigm with integrated SiPh. Starting with the quantum source part, generally, two types of solid-state implementations may be applied for the quantum sources serving single photons: (1) on-demand single-photon generation from single quantum dots, defects, or impurities [18,19], weakly coupled to optical microcavities or nanocavities [20], and (2) heralded single-photon generation from on-chip optical waveguides through nonlinear optical processes such as spontaneous parametric down-conversion (SPDC) [21] or spontaneous four-wave mixing (SFWM) [22-25]. To produce single photons that are highly indistinguishable and pure, the approach of applying quantum dots, defects, or impurities requires cooling the systems down to cryogenic temperatures typically < 4 K to avoid phonon broadening of spectral lines, and therefore may not be suitable for room-temperature PQC. On the other hand, with the advancement of spatial, temporal, and spectral multiplexing techniques to overcome the probabilistic nature of heralded single-photon generation [22-24], quantum sources that generate single photons through on-chip optical waveguides with off-resonant optical nonlinearities have becoming practical for room-temperature PQC. Here, since SOI waveguides commonly used in SiPh inherently possess the $3^{rd}$ order Kerr nonlinearity, SFWM source is assumed in the proposed room-temperature PQC paradigm with SiPh. Moreover, an increased value of Kerr index $n_2$ can be attained if SiGe alloy is employed as the ring-waveguide material in the SFWM sources.

For the part of quantum circuits, we assume the use of a field-programmable interferometer mesh (FPIM) implemented by cascaded Mach-Zehnder interferometers (MZIs) [26,27], so that various quantum algorithms can be performed through manipulating the path degree of freedom of single photons at room temperature. Note that although in Fig. 1 the quantum circuits are composed of SOI waveguides, lower-loss silicon nitride (SiN) waveguides can also be used. Finally, for the part of quantum detectors, SPDs and NPDs based upon the proposed waveguide GeSi SPADs operated at room temperature are assumed, which will be the main topic of discussion in the following sections.

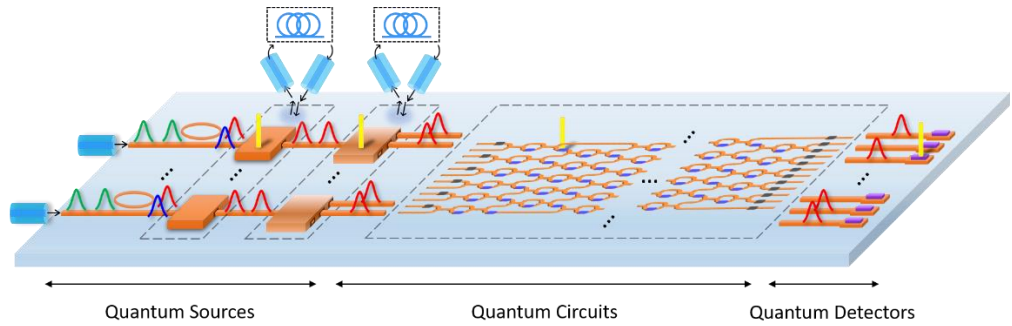

**Fig. 1.** Schematic plot of the proposed room-temperature PQC paradigm using the path degree of freedom of single photons: single photons are generated through SFWM (green pulses converted to blue and red pulses) in SOI rings (orange circles) followed by filters (orange box that blocks the blue pulses) and active temporal/spatial multiplexers (orange box that converts serial pulses to parallel pulses) (quantum sources), manipulated by a FPIM using cascaded MZIs (quantum circuits), and measured by the proposed waveguide GeSi SPADs as SPDs and/or NPDs (quantum detectors). An application-specific integrated circuit (ASIC) layer is assumed to be flipped and bonded on the PIC layer with copper (Cu)-Cu pillars (yellow lines) connected wafer-level hybrid bond, or with metal bumps (yellow lines) connected chip-on-wafer-on-substrate (CoWoS) packaging. The off-chip fiber couplings are either for the pump lasers or the optical delay lines.

## 3. Waveguide GeSi SPAD

### 3.1 Structural Design

Translating the design of an avalanche photodiode (APD) from normal incidence geometry to waveguide geometry is less straightforward compared to the case of a photodiode (PD). This is due to the additional carrier multiplication layer that is typically much thicker than the SOI layer thickness, making an efficient coupling between the single-mode SOI waveguide and the waveguide GeSi APD hard to achieve. As an example, even for the low-breakdown-voltage design ~ -10 V demonstrated in Ref. [9], the thickness of the Si multiplication layer can be as thick as 670 nm and is not compatible with the standard 220 nm SOI layer thickness. An inverse taper connecting the SOI waveguide to the Si multiplication layer is one possible solution. However, the evanescent coupling between the Ge absorption layer and the Si multiplication layer is rather weak because of the effective index mismatch especially for thicker waveguiding layers. It is possible to create a butt coupling between the end of the inverse taper and the beginning of the Ge absorption layer, but the fabrication processes may be much more complicated. Here we consider a step coupler [28] that connects the SOI waveguide and the Si multiplication layer, and, due to the multi-mode interference (MMI) in the vertical direction, the optical fields oscillate up-and-down in the waveguide GeSi APD so that Ge absorption layer absorbs light efficiently. In addition, an aluminum (Al) back mirror at the end of the device is considered to induce reflections that effectively increase the absorption length so that the actual device length can be further minimized. See Fig. 2 for the schematic plots of the proposed waveguide GeSi APD.

The structure shown in Fig. 2 may be fabricated by either a top-down or a bottom-up approach. For the top-down approach, blanket epitaxy of the Si multiplication layer and the Ge absorption layer is first applied, followed by a series of dry and wet etches to sequentially define the Ge mesa, Si mesa and SOI waveguides. In this case, special attention should be paid to the surface roughness of the SOI waveguides since both their top surface and the sidewalls are formed by etching, which may be challenging to achieve ultralow waveguide loss. Alternatively, as previously mentioned, SiN waveguide may be applied instead and fabricated right before the Al back mirror fabrication to serve as the main waveguiding backbone. For the bottom-up approach, selective epitaxy of the Si multiplication layer in a patterned oxide trench is applied, followed by a chemical-mechanical polishing (CMP) to form the Si mesa. Then, similar processes are repeated to form the Ge mesa. In this case, complex fabrication steps such as selective epitaxy and mesa CMP are involved so it may be challenging to determine the recipes for selective epitaxy and mesa CMP that respectively maintain the high material quality and planarize the mushroom-shape like Si and Ge mesas effectively. In both approaches, the Al back mirror is fabricated by first forming a trench right after the end of the waveguide GeSi SPAD by etching oxide, followed by depositing and patterning Al to serve as the back mirror. Afterwards, the large topography may be challenging to be filled by oxide deposition and planarized by oxide CMP before finally going through the backend metal processes.

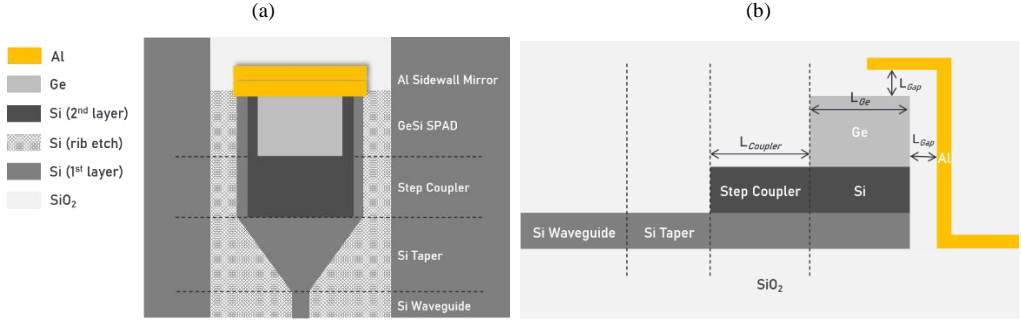

**Fig. 2.** (a) Top view of the proposed waveguide GeSi SPAD, in which the materials assumed are listed. (b) Cross-sectional view of the proposed waveguide GeSi SPAD, in which the variables for optimizing QE are illustrated.

### 3.2 QE optimization

Next, finite-difference time domain (FDTD) simulations are performed to optimize the QE of the proposed waveguide GeSi SPAD. Following Ref. [9], the Ge absorption layer and the Si multiplication layer thicknesses are set to be 450 nm and 670 nm, respectively. The SOI layer thickness is assumed to be 220 nm and the buried oxide (BOX) thickness is assumed to be 2 µm. Due to the tensile strain induced by the thermal expansion coefficient mismatch between Ge and Si during the epitaxy process, to accurately account for the Ge absorption coefficients around the C bands, experimental data in Ref. [9,29] are used in the FDTD simulations. The standard simulation procedure to calculate the QE is to 1) define the material n/k values, the device geometry, and the operation wavelength, 2) excite the SOI waveguide with its fundamental mode, 3) monitor the absorption in the Ge region.

Fig. 3(a) shows the simulated QE at 1550 nm wavelength as a function of step coupler length ($L_{Coupler}$) and Ge absorber length ($L_{Ge}$), without the addition of the Al back mirror. The beat lengths of the MMI can be clearly observed at 1.4 µm, 5.3 µm, and 9.2 µm, and the highest QE achievable for Ge length < 15 µm is about 90 %. Fig. 3(b) shows the simulated QE at 1550 nm wavelength as a function of gap length ($L_{Gap}$) and Ge length ($L_{Ge}$), with the addition of the Al back mirror. It is observed that QE larger than 95 % can be obtained when the coupler length, the gap length, and the Ge length are appropriately chosen. In Fig. 3(c), the spectral QEs centered at 1550 nm and 1310 nm wavelengths are plotted, and for both cases the QEs can be larger than 95 % at the two center wavelengths. The rapid decrease of QEs at wavelengths > 1550 nm is due to the rapid decrease of absorption coefficients when the photon energy becoming smaller than the Ge Γ valley energy bandgap. Note that while not shown here, the QEs at 1310 nm are scanned in similar manners as in Fig. 3(a) and 3(b), so that when picking the final values for parameters $L_{Coupler}$, $L_{Ge}$, and $L_{Gap}$, the QEs at the two center wavelengths can be examined and maximized at the same time.

(a)     (b)

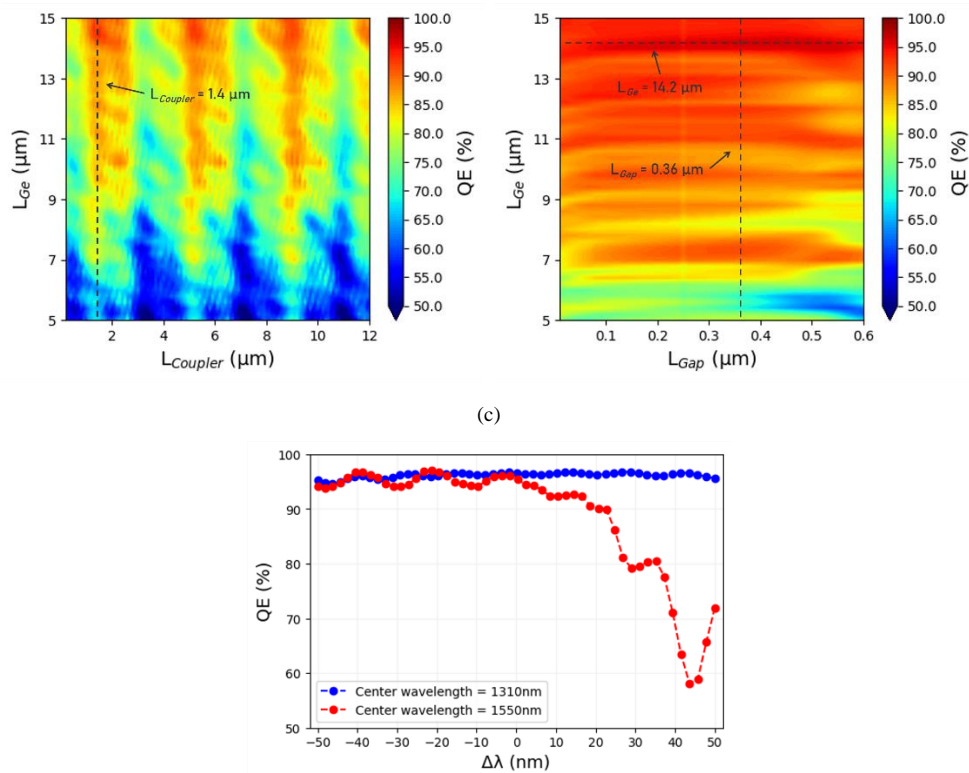

(c)

**Fig. 3.** (a) QE of the proposed waveguide GeSi SPAD without the Al back mirror, simulated at 1550 nm as a function of coupler length and Ge length. (b) QE of the proposed waveguide GeSi SPAD with the Al back mirror, simulated at 1550 nm as a function of gap length and Ge length. (c) QE of the proposed waveguide GeSi SPAD with the Al back mirror, simulated as a function of wavelength centered at 1550 ± 50 nm (around the C band) and at 1310 ± 50 nm (around the O band), given the optimal conditions, i.e., coupler length equal to 1.4 µm, gap length equal to 0.36 µm, and Ge length equal to 14.2 µm. While the above data are obtained by 2D FDTD simulations, we also verify that for Ge width > 1 µm and mesa design rule < 200 nm, there is little difference between the data obtained by 2D and 3D FDTD simulations.

## 3.3 DCR scaling

In Ref. [9], at room temperature, a normal-incidence GeSi SPAD with 15 µm active region diameter exhibits DCRs as low as 5 MHz and 10 MHz at excess biases of 1 V and 2 V, respectively, along with a low breakdown voltage at -10.26 V. Note that the active region diameter is defined as the diameter of the Ge mesa. For the proposed waveguide GeSi SPAD, the estimated sizes of Ge mesa would be 14.2 µm long based on the simulations in Sec. 3.2, and 2 µm wide so that there is sufficient surface area to place contact vias without perturbing the light in the Ge absorption layer. To estimate the DCR more accurately when scaling the sizes of the Ge mesa, we refer to the dark current data measured over five reference PDs different in diameters [9,30]. These reference PDs are originally designed to locate the unity gain point of the corresponding APDs, but can also be used to extract the device bulk and surface dark current densities. In Fig. 4, the measured dark currents at -1 V reverse bias normalized by the Ge mesa perimeter are plotted as a function of Ge mesa diameter. In this representation, the intercept and the slope of the curve correspond to the surface dark current density and bulk dark current density, and they are found to be 4.12 µA/cm$^2$ and 0.7 nA/cm, respectively. These are the lowest values reported in the literature to the best of our knowledge, which enable the room-temperature operation of GeSi SPAD reported in Ref. [9]. Based on the above numbers on areas and dark current densities, the DCRs of the proposed waveguide GeSi SPAD are estimated to be ~ 0.8 MHz and 1.6 MHz at excess biases of 1 V and 2 V, respectively.

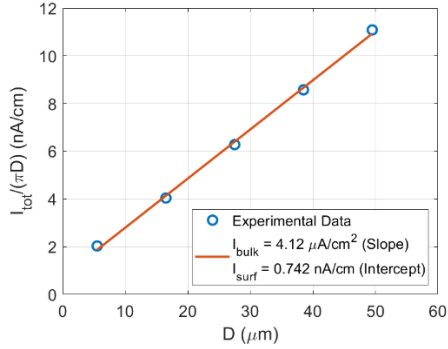

**Fig. 4.** Dark current of GeSi PD at -1 V reverse bias, normalized by its active region circumference, plotted as a function of active region diameter. The experimental data (blue dots) consist of the average dark current between two device repeats (the ratio of the standard deviation to the average is < 2%) for five different active region diameters. The linear fitting (red line) shows the bulk dark current density and the surface dark current density with its slope and intercept, respectively.

## 4. Benchmarks between GeSi SPAD and SNSPD for PQC

### 4.1 Considerations on figures of merit of SPD and NPD for PQC

To evaluate the performance of SPDs, noise-equivalent power (NEP) that is proportional to the square root of the DCR and inversely proportional to the single-photon detection efficiency (SPDE), has been commonly used. This is a signal-to-noise-ratio like figure of merit, and is useful for applications when the lowest optical power of a continuous wave (CW) or a pulse that can be detected is concerned. For example, in light detection and ranging (LiDAR) application, it is often desired to know the maximum distance that a LiDAR system can perceive, which is directly related to the NEP of SPDs. However, it has been pointed out that in quantum key distribution (QKD) application, the contribution to quantum-bit error rate by SPDs is directly related to the ratio between the DCR to the sifted photon count rate (PCR) [31], so that a signal-to-background-ratio like figure of merit, e.g., DCR multiplied by timing jitter (TJ) divided by SPDE, should be used. Here the timing jitter refers to the minimal time-gating window that can be applied. Therefore, choosing the right figure of merit highly depends on the key metrics of interest in different applications.

For the quantum detectors in PQC application, the key metric should be whether a photon state can be faithfully detected in the presence of the DCR for each run of the experiment. In this sense, signal-to-noise-ratio or signal-to-background-ratio like definitions not only are loosely related to this metric, but also put too much emphasis on the DCR due to its placement in the denominator of these ratios. In the following, we derive several figures of merit that are suitable for the quantum detectors used in PQC in general.

### 4.2 Derivation of figures of merit of SPD and NPD for PQC

First, we assume all photon detection events occur in a single or multiple pre-defined time-gating windows with a temporal width $T_G >$ TJ. While this is not a valid assumption for LiDAR application because it is not possible to know the time-of-flight of all photons reflected from a scene in advance, this is a reasonable assumption for PQC application because the time-of-flight of single photons in pre-defined quantum circuits can be predictable. This is especially true for the FPIM assumed in Fig. 1, in which single photons reach the quantum detectors at multiple instances separated by a constant temporal period that is associated with the delay introduced by an individual MZI. In summary, as to be shown in the following, the time-gating largely eliminates the deleterious effects of the SPAD DCR in the proposed room-temperature PQC paradigm.

For the scheme of photon-based PQC, we assume each of the output ports of the FPIM is terminated with an NPD consisting of a spatially-multiplexed array of waveguide SPDs. When an $N$ photon state is measured, the "click" and "success" probabilities of photon detection can be defined as

$$P_{click} = P_{OO}\left(P_{PC}(1-P_{DC}) + (1-P_{PC})P_{DC} + P_{PC}P_{DC}\right)^N (1-P_{DC})^{M-N}$$
$$P_{success} = P_{OO}(P_{PC})^N (1-P_{DC})^{M-N}$$

(1)

$P_{DC}$ and $P_{PC}$ are the probabilities that the "click" is due to DCR and PCR, respectively. $M$ is the number of waveguides for SPDs, and $N$ is the number of photons into an NPD. $P_{OO}$ is the probability corresponding to the "one-to-one" assignment of a single photon to a waveguide SPD, and can be derived as

$$P_{OO} \underset{M \geq N}{=} \frac{\frac{1}{N!}\prod_{n=1}^{N}(M+1-n)}{C_N^{M+N-1}} = \prod_{n=1}^{N} \frac{M+1-n}{M+N-n}.$$

(2)

Assuming the dark counts arrive with Poisson distribution and the photo counts arrive deterministically, $P_{DC}$ and $P_{PC}$ can be further re-written as

$$P_{DC} = 1 - e^{-R_{DC}T_G}$$
$$P_{PC} = \eta_{SPD}$$

(3)

where $R_{DC}$ is the DCR and $\eta_{SPD}$ is the SPDE. Finally, we define the "fidelity" of photon detection by the ratio between the "success" probability and the "click" probability, and it can be derived as

$$F = \frac{P_{success}}{P_{click}} \underset{NR_{DC}T_g \ll 1}{\sim} 1 - NR_{DC}T_G \left(\frac{1-\eta_{SPD}}{\eta_{SPD}}\right),$$

(4)

given a good $\eta_{SPD}$ and negligible $NR_{DC}T_G$. Note that SPD corresponds to $M=1$ and $N=1$ whereas NPD corresponds to $M \geq N$ and $N \geq 1$. Experimentally, we consider two types of spatial multiplexing schemes to assemble an NPD from an array of SPDs. The first type in shown in Fig. 5(a), in which the input SOI waveguide is butt coupled to a star coupler and then directed into an array of waveguide SPDs. The length and width of the star coupler are estimated to be 78 µm at 1550 nm (92 µm at 1310 nm) and 28 µm, respectively, when the input/output waveguides are tapered to/from 1 µm wide and the output waveguide separation is tapered from 200 nm to cover an array of 20 waveguide SPDs. Note that similar concept for SNSPDs to be photon-number-resolving has been discussed [32], and the areas of the GeSi SPAD version and the SNSPD version should be similar. The second type is shown in Fig. 5(b), in which the input SOI waveguide is evanescently coupled to a side waveguide and then directed into an array of waveguide SPDs. The length and width of the side waveguide are estimated to be 2 µm at 1550 nm (3 µm at 1310 nm) and 0.5 µm, respectively, when the input waveguide is 110 nm (120 nm) away from the side waveguide to cover an array of 20 waveguide SPDs. Again, note that similar concept for SNSPDs to be photon-number-resolving has been discussed [33], and the area of the GeSi SPAD version area may be larger than that of the SNSPD version mainly because the superconducting niobium (Nb) can be used for optical absorption and electrical conduction as the same time.

(a) (b)

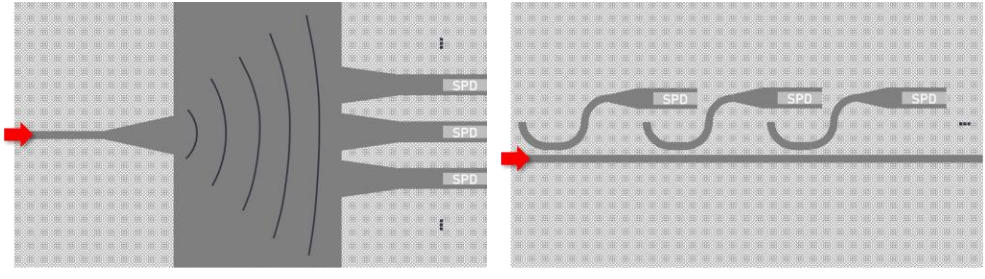

**Fig. 5.** Possible configurations of a waveguide NPD based on a spatially-multiplexed array of waveguide SPDs. The input SM waveguide is coupled to an array of waveguide SPDs through (a) a star coupler or (b) cascaded evanescent waveguide couplers.

For the scheme of qubit-based PQC, we assume the dual-rail representation for the path degree of freedom of a single photon [34], and each of the output ports of the FPIM is terminated with an SPD. When an $N$ qubit state is measured, the "click" and "success" probabilities of qubit detection can be defined as

$$P_{click} = \left(P_{PC}(1-P_{DC}) + (1-P_{PC})P_{DC} + P_{PC}P_{DC}\right)^N (1-P_{DC})^N$$
$$P_{success} = P_{PC}^N (1-P_{DC})^N$$
(5)

Similarly, the "fidelity" of qubit detection can be defined by the ratio between the "success" probability and the "click" probability, and it can be derived as

$$F = \frac{P_{success}}{P_{click}} \underset{NR_{DC}T_g \ll 1}{\sim} 1 - NR_{DC}T_G\left(\frac{1-\eta_{SPD}}{\eta_{SPD}}\right),$$
(6)

given a good $\eta_{SPD}$ and negligible $NR_{DC}T_G$. Note that Eq. (6) is identical to Eq. (4). In the following, we shall apply these figures of merit to compare the performance between GeSi SPAD and SNSPD for PQC.

### 4.3 Comparison between GeSi SPAD and SNSPD for PQC

The following parameter values are used for comparing GeSi SPAD at 300 K and SNSPD at 4 K. For the case of GeSi SPAD at 300 K, SPDE is assumed to be 95 % and DCR is assumed to be 1.6 MHz based on the above simulations and calculations, considering 100 % avalanche trigger probability at 2 V excess bias. For the case of SNSPD at 4 K, SPDE is assumed to be 91 % and DCR is assumed to be 5886 Hz based on Ref. 12. In both cases $T_G$ is assumed to be 1 ns, which can be easily generated by high-speed electronics and larger than the TJs of GeSi SPAD at 300 K and SNSPD at 4 K that are usually between tens to hundreds of picoseconds.

For the scheme of photon-based PQC, in Fig. 6(a) and 6(b), the success probability and the fidelity of photon detection considering 300 K GeSi SPAD are respectively plotted as a function of $M$ waveguides and $N$ photons. In Fig. 6(c) and 6(d), the difference in the success probabilities and the difference in the fidelities of photon detection considering 300 K GeSi SPAD and 4 K SNSPD are respectively plotted as a function of $M$ waveguides and $N$ photons. By inspecting 6(c) and Fig. 6(d), it is observed that the success probability of 300 K GeSi SPAD is a few percents (e.g., 4 % when $N$=1) higher than that of 4 K SNSPD due to the assumed SPDEs, and the fidelity of 300 K GeSi SPAD is 0.0084 % lower than that of 4 K SNSPD. Note that it is expected that the success probabilities shown in Fig. 6(a) go up when $M$ is much larger than $N$, due to the reduced probability of finding more than one excitation in each waveguide SPD. For the scheme of qubit-based PQC, in Fig. 7(a), the success probability and the fidelity of qubit detection considering 300 K GeSi SPAD are respectively plotted as a function of N qubits. In Fig. 7(b), the difference in the success probabilities and the difference in the fidelities of qubit detection considering 300 K GeSi SPAD and 4 K SNSPD are respectively plotted as a function of N qubits. By inspecting Fig. 7(b), it is observed that the success probability of 300 K GeSi SPAD is a few percents (e.g., 3.848 % when N=1) higher than that of 4 K SNSPD due to the assumed SPDEs, and the fidelity of 300 K GeSi SPAD is again 0.0084 % lower than that of 4 K SNSPD.

These results showcase that for PQC, a SPDE as high as possible plays the key role but an ultralow DCR is not a must. Conventional figures of merit based on signal-to-noise-ratio or signal-to-background-ratio like definitions do not fully capture the needs of PQC, and the proposed waveguide GeSi SPAD at 300 K can perform as well as a SNSPD at 4 K when considering new figures of merit suitable for PQC application as discussed in this paper. Moreover, since room-temperature operation of normal-incidence GeSi SPAD fabricated by CMOS compatible processes has been recently demonstrated [9], we expect that waveguide GeSi SPAD shall soon be implemented experimentally.

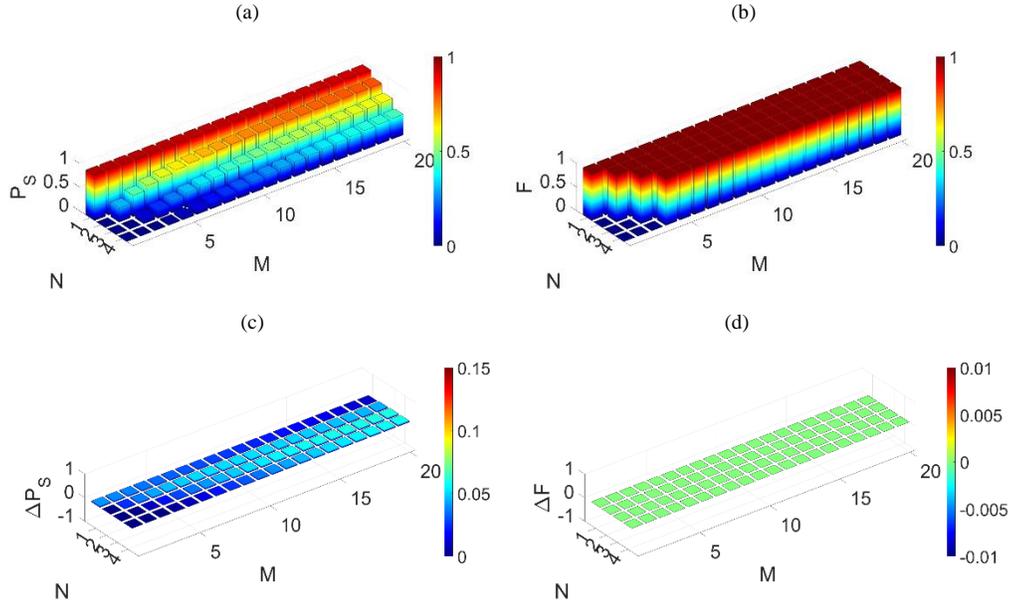

**Fig. 6.** For the scheme of photon-based PQC: (a) The probability of successfully detecting N photon state, and (b) the fidelity of detecting N photon state, using M spatially-multiplexed waveguide GeSi SPADs at 300 K as an NPD. (c) The difference in the probabilities of successfully detecting N photon state, and (b) the difference in the fidelities of detecting N photon state, using M spatially-multiplied waveguide GeSi SPADs at 300 K and SNSPDs at 4 K as NPDs. Note that no approximation is used in the formula for plotting these figures.

(a)            (b)

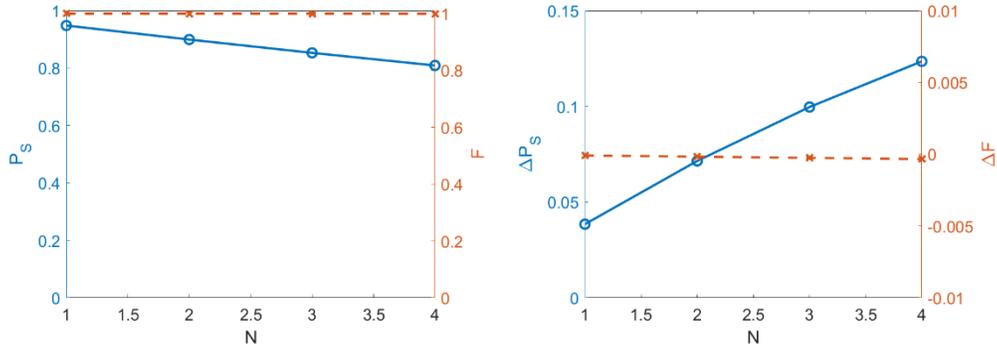

**Fig. 7.** For the scheme of qubit-based PQC: (a) The probability of successfully detecting $N$ qubit state, and the fidelity of detecting $N$ qubit state, using single waveguide GeSi SPADs at 300 K as SPDs. (b) The difference in the probabilities of successfully detecting $N$ qubit state, and the difference in the fidelities of detecting $N$ qubit state, using single waveguide GeSi SPADs at 300 K and SNSPDs at 4 K as SPDs. Note that no approximation is used in the formula for plotting these figures.

## 5. Discussion and Conclusion

So far, the operation wavelength of the proposed room-temperature PQC paradigm is assumed to be at SWIR. It is known that two-photon absorption (TPA) in SOI waveguide at SWIR may affect the heralding efficiency of a SFWM source. In this context, MIR based quantum optics experiments [13] has been proposed and demonstrated, but the lack of appropriate SPD and NPD hinders further progress. SPADs based on GeSn on Si [14], GeSn on Ge on Si [15,16], and strained superlattice on Ge on Si [17] may be feasible solutions, but further investigations on the material quality are needed to evaluate their applicability for room-temperature operation.

To conclude, we propose and analyze a room-temperature-operated waveguide GeSi SPAD on the integrated SiPh platform based on the recently demonstrated normal-incidence GeSi SPAD at room temperature. We show that the proposed device with a reasonable time-gating window is the key enabler for PQC systems to be operated at room temperature, with detailed performance parameters and suitable figures of merit. Our findings pave a new path toward cryogenics-free PQC that shall significantly increase the testing throughputs and so reduce the design iteration cycles, which are essential for PQC to continuously evolve to full fruition.

## 6. Supplementary Material

The source codes for designing and simulating Fig. 3(a), (b), and (c) with commercial FDTD software from Lumerical are included in the Supplementary Material.

**Funding.** The research of RS is supported by the U.S. Air Force Office of Scientific Research on grant FA9550-21-1-0347.

**Acknowledgments.** NN would like to thank Dr. Y.-C. Lu and Mr. Y.-H. Liu for discussing the dark current data.

**Disclosures.** NN, CYH, and EC are shareholders of Artilux Inc., a private company that makes SWIR 2D/3D sensors, imagers, and photonic integrated circuits.

**Data availability.** The data supporting this work are available in the main text, figures and Supplementary Materials. Further information is available from the corresponding author.